\documentclass[conference]{IEEEtran}
\IEEEoverridecommandlockouts
\usepackage{balance}
\usepackage{cite}
\usepackage{amsmath,amssymb,amsfonts}
\usepackage{algorithmic}
\usepackage{graphicx}
\usepackage{textcomp}
\usepackage{xcolor}
\usepackage{subfig}
\newtheorem{Lemma}{Lemma}
\newtheorem{Corollary}{Corollary}
\def\BibTeX{{\rm B\kern-.05em{\sc i\kern-.025em b}\kern-.08em
    T\kern-.1667em\lower.7ex\hbox{E}\kern-.125emX}}
\begin{document}

\title{SoReC: A Social-Relation Based Centrality Measure in Mobile Social Networks
}
\author{
\IEEEauthorblockN{Bowen Li\IEEEauthorrefmark{1}\IEEEauthorrefmark{2}, Zhenxiang Gao\IEEEauthorrefmark{3}, Xu Shan\IEEEauthorrefmark{1}, Weihua Zhou\IEEEauthorrefmark{1}\IEEEauthorrefmark{4}, Emilio Ferrara \IEEEauthorrefmark{3}}

\IEEEauthorblockA{\IEEEauthorrefmark{1}Institute of Information Engineering, Chinese Academy of Sciences, Beijing 100093, China}
\IEEEauthorblockA{\IEEEauthorrefmark{2}School of Cyber Security, University of Chinese Academy of Sciences, Beijing 100049, China}
\IEEEauthorblockA{\IEEEauthorrefmark{3}Information Sciences Institute, University of Southern California, CA 90292, USA}
\IEEEauthorblockA{\IEEEauthorrefmark{4}Corresponding author}
E-mail: libowen@iie.ac.cn, zhenxiang.gao.mail@gmail.com, shanxu@iie.ac.cn, zhouweihua@iie.ac.cn, emiliofe@usc.edu
}
\maketitle

\begin{abstract}
Mobile Social Networks (MSNs) have been evolving and enabling various fields in recent years. Recent advances in mobile edge computing, caching, and device-to-device communications, can have significant impacts on 5G systems. In those settings, identifying central users is crucial. It can provide important insights into designing and deploying diverse services and applications. However, it is challenging to evaluate the centrality of nodes in MSNs with dynamic environments. In this paper, we propose a Social-Relation based Centrality (SoReC) measure, in which social network information is used to quantify the influence of each user in MSNs. We first introduce a new metric to estimate direct social relations among users via direct contacts, and then extend the metric to explore indirect social relations among users bridging to their neighbors. Based on direct and indirect social relations, we detect the influence spheres of users and quantify their influence in the networks. Simulations on real-world networks show that the proposed measure can perform well in identifying  future influential users in MSNs.

\end{abstract}


\section{Introduction}\label{Section1}
The advances in popularity of wireless networks and mobile devices bring unprecedented prosperity to Mobile Social Networks (MSNs) \cite{Vastardis2013Mobile, gao2015exploiting}. Millions of mobile users can directly connect, interact, and share content with each other via their smart devices, which become one of the most important paradigms in the 5G system \cite{Wang2018D2D}. In this paradigm, centrality evaluation is a key research issue \cite{Chai2013Cache, Barbera2014Data, Kim2012Temporal, Li2009LocalCom, Zhou2017Predicting}. It is helpful to identify the influential users in the networks, as this provides important insights into the design and deployment of diverse applications and services in various settings, such as mobile edge computing, content-centric networks, device-to-device communication, etc.
Most research \cite{Chai2013Cache, Barbera2014Data} on centrality evaluation in MSNs is based on the \textit{static-network assumption}. The topologies of the networks are supposed to remain the same over time, e.g., a link between two users exists if they had interacted within the observation period, and doesn't exist otherwise. In other words, any temporal information is essentially disregarded. However, MSNs consisting of mobile devices carried by humans are essentially dynamic environments, i.e., link vary over time, often significantly. Thus, even quite effective centrality measures for static networks, e.g., degree, closeness, node and edge betweenness \cite{Freeman1978Centrality, de2012novel}, and PageRank \cite{Sehgal2009The}, are not ideal for dynamic MSNs. In order to evaluate the centrality of users in such challenging network environments, some researchers \cite{Kim2012Temporal, gao2015measures} built a time-ordered model according to human mobility patterns and tried to quantify the influence of each user by capturing the spatial and temporal characteristics of the networks. While many researchers have studied that mechanism as a means of centrality evaluation in MSNs, the effects of the social nature of MSNs have generally been ignored when considering how to predict the centrality of users over time. Further studies \cite{Gonz2008Understanding, Eagle2009Inferring} show that spatial and temporal actions (mobility) of humans are not chaotic but are strongly impacted by social relations --- these social relations, in turn, have stable long-term characteristics. Thus, social relations need to be taken into consideration when evaluating centrality. In addition, apart from the direct relations, indirect relations also need to be considered, because even two nodes with no direct relation can still have a strong influence on each other as long as the two have some mutual friend(s) --- indirect connections can drive social influence. The importance of developing accurate centrality measures in MSNs is further enhanced by the fact that such centrality measures are also often exploited in downstream tasks such as community detection and recommendation systems \cite{de2013enhancing, de2014mixing}.

In this paper, we investigate the centrality evaluation from the perspective of social relations and propose a centrality measure to identify influential users in dynamic MSNs. First, direct and indirect social relations are studied. A new metric is proposed to estimate direct social relations among contact users by mining their contact patterns. We also give a brief (mathematical and experimental) proof of the metric validity. Apart from the direct social relations, indirect social relations are studied to estimate the relations among users bridging to their neighbors. Combing direct and indirect social relations, we propose a \textit{Social-Relation based Centrality} (SoReC) measure to quantify the centrality of users in dynamic MSNs. Extensive simulations on real-world mobility networks show that the SoReC measure can well identify future influential users in MSNs.
The rest of the paper is organized as follows: In Section \ref{section2}, we briefly depict the mobile social network model. In Section \ref{section3}, we detail our framework for centrality evaluation. In Section \ref{section4}, we conduct the performance evaluation and discuss the results. Finally, we conclude the paper along with insights into future directions in Section \ref{section5}.

\section{Network Model}\label{section2}

Consider a mobile social network, which consists of $N$ mobile devices. Each mobile device can directly communicate with others over short-range radio frequencies when they are within the direct transmission range of each other. For each time slot $t$, the transient MSN is static and denoted as an undirected unweighted graph $G_{t} = \left( V_{t}, E_{t}\right)$, where $V_t$ is a set of nodes representing all mobile devices in the network at the time slot $t$, $V_t=\{v_i\}$, $1 \leq i \leq N$, and $E_t$ is a set of edges representing the interaction states among the mobile devices at the time slot $t$, $E_{t}= \{\left(v_i,v_j\right)|\,d(v_i,v_j) \leq D,\,v_i \in {V}_{t},\,v_j \in {V}_{t}\}$, where $d(v_i,v_j)$ denotes the physical distance between nodes $v_i$ and $v_j$. When $d(v_i,v_j)$ is not less than a special distance $D$ (the maximum wireless transmission distance), the direct interaction between them can occur, thereby an edge $\left(v_i,v_j\right)$ forms, otherwise not. If $(v_i,v_j) \in E_{t}$, we say $v_i$ and $v_j$ are adjacent.
We assume the time during which the network is observed is finite, from $t_\text{start}$ until $t_\text{end}$; Without loss of generality, we set $t_\text{start} = 0$ and $t_\text{end} = T$. The dynamic MSN in the time interval $[0,T]$ is expressed as a time-ordered network $\mathcal{G} = \{G_{0},\dots,G_{T} \}$.

\section{Framework of Centrality Evaluation}\label{section3}

This section details our framework of centrality evaluation, which comprises of two parts: \textit{(i)} the social relations analysis, and \textit{(ii)} the centrality quantification. In the former, direct and indirect social relations are explored, respectively. In the latter, the concept of influence spheres is introduced first and then a centrality measure is proposed.

\subsection{Analysis of Direct Social Relations}\label{section3A}

The transient nature of the connectivity among nodes, which enables messages to travel over the MSNs, yields challenges in detecting the interaction probabilities among nodes. Since direct interactions among nodes only occur when nodes come into the wireless transmission range of each other, direct social relations arising from physical proximity (contact) need to be analyzed to evaluate interaction intensity (influence strength).

\begin{figure}[t]
\centerline{\includegraphics[width=0.48\textwidth]{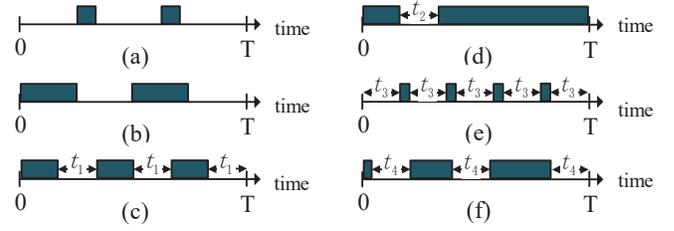}}
\caption{Six different contact patterns between nodes $v_i$ and $v_j$ during $[0, T]$, where shaded boxes represent the encounter duration between them.}
\label{fig1}
\end{figure}

Previous studies have proposed diverse metrics to extract the intensity of direct social relations, such as encounter frequency (EF), total contact duration (TCD), and average separation period (ASP) \cite{Li2009LocalCom,Zhou2017Predicting}. But all those metrics have some inadequacies in reflecting the interaction intensity arising from the contacts. For example, consider the six contact patterns in Fig.~\ref{fig1}, where the shaded boxes represent the contacts' duration. Comparing case (a) with case (b), we notice that $EF(a) = EF(b)$ but $TCD(a) < TCD(b)$. Hence, the contact pattern (b) captures a stronger interaction. In cases (b) and (c), $TCD(b) = TCD(c)$ but $EF(b) < EF(c)$. Since frequent encounters bring more interactions, pattern (c) is preferable.

Among the previous metrics, the metric $EF$ cannot differentiate between case (a) and case (b), and the $TCD$ cannot differentiate between case (b) and case (c). Although $ASP$ can assign correct link weights in cases (a), (b) and (c), it fails in other cases. For example, consider case (c) and case (d). If $t_1 = t_2$, $ASP$ cannot differentiate between them but case (d) is preferable due to the longer uninterrupted contact duration. Similarly, if $t_2 = t_3$, $ASP$ cannot differentiate between case (d) and case (e), even though case (d) offers better content transmission opportunities. Meanwhile, the variance of the contact time is also a factor reflecting the irregularity in the relationship, but all the three metrics are unable to reflect it. Such as for case (c) and case (f), if $t_3 = t_4$, then $EF(c) = EF(f)$, $TCD(c) = TCD(f)$, $ASP(c) = ASP(f)$. Since a more stable encounter duration captures a more stable content transmission, the relationship in case (c) is preferable to interaction opportunity.

To find a metric that reflects the direct social relation more accurately, we propose a new metric, \textit{Social-Relation Stability} (SRS), by taking into consideration the aforementioned three factors: frequency, duration, and regularity. The idea is as follows: calculate the contribution of each encounter by the sine function and add them up. We denote $\Theta_{v_i,v_j}$ as the contact patterns between nodes $v_i$ and $v_j$ during $[0,T]$, $\Theta_{v_i,v_j} = \{\theta_1,\dots \theta_K\}$ ($\sum \theta_k \le T$), where $\theta_k$ is the duration of the $k$-th encounter. Hence the SRS metric is defined as
\begin{equation}
{SRS_{v_i,v_j}} = \frac{\sum_{k = 1}^K {f\left( {{\theta_k}} \right)}}{\pi / 2},
\label{eq3}
\end{equation}
where $f\left( {{\theta_k}} \right)  =\sin(\pi \theta_k/2 T)$. Note that, since the sine function is monotonically increasing and concave in the partial interval, the value of SRS is positively correlated with the frequency, the longevity, and the regularity of interactions. Next, we will mathematically prove the validity of the SRS metric (in Lemma \ref{lemma1}-\ref{lemma3}) and discuss the range of metric values (in Corollary \ref{Corollary}).

\begin{Lemma}\label{lemma1}
The SRS value is positively correlated with contact frequency.
\end{Lemma}

\begin{IEEEproof}
Without loss of generality, we assume the total contact duration is constant and contacts' duration are regular. Substituting $f\left(\theta_k\right)$ into (\ref{eq3}), we can get
\begin{equation}
{SRS_{v_i,v_j}} = \frac{2}{\pi} \cdot \sum\limits_{k = 1}^K {\sin \left( {\frac{\pi }{2} \cdot \frac{\theta_k}{T}} \right)}.
\label{eq4}
\end{equation}
Due to the regular contacts' duration, i.e., $\forall \theta_k, \theta_t \in \Theta_{v_i,v_j}, \theta_k = \theta_t$, (\ref{eq4}) can be rewritten as
\begin{equation}
{SRS_{v_i,v_j}} = \frac{2}{\pi} \cdot K \cdot \sin \left( { \frac{\pi }{2} \cdot \frac{T_{meet}/K}{T} }\right),
\label{eq5}
\end{equation}
where $T_\text{meet}$ denotes the total contact duration, $T_\text{meet} = \sum_{\theta_k \in \Theta_{v_i,v_j}} \theta_k $. Since ${\partial {SRS_{v_i,v_j}}} / {\partial K} > 0$ when $K \geq 1$ (${SRS_{v_i,v_j}}=0$, when $K =0$), $SRS_{v_i,v_j}$ increases with the increasing of $K$. Thus, the SRS value is positively correlated with contact frequency.
\end{IEEEproof}

\begin{Lemma}\label{lemma2}
The SRS value is positively correlated with contact duration.
\end{Lemma}

\begin{IEEEproof}
Similarly, we assume the contact frequency is constant and contacts' duration are regular. Due to the regular contacts' duration, the SRS metric can be also derived as (\ref{eq5}). Since ${\partial {{SRS_{v_i,v_j}}}} /  {\partial {T_\text{meet}}} > 0$ within $\left[ {0, \space T}\right]$ and the encounter frequency $K$ is a constant, $SRS_{v_i,v_j}$ increases with the increasing of $T_\text{meet}$. Thus, the SRS value is positively correlated with contact duration.
\end{IEEEproof}

\begin{Lemma}\label{lemma3}
The SRS value is positively correlated with contact regularity.
\end{Lemma}

\begin{IEEEproof}
For each $\theta_k \in \Theta_{i,j}$, $0 \leq \theta_k \leq T$, i.e., $\pi \theta_k / 2T \in [0,\pi/2]$, so $f\left(\theta_k\right)$ is concave. According to the Jensen Inequality, we can obtain
\begin{equation}
f \left({\frac {\sum a_{k}\theta_{k}}{\sum a_{k}}}\right)\geq {\frac {\sum a_{k} f (\theta_{k})}{\sum a_{k}}},
\label{eq6}
\end{equation}
where $a_{k}$ is the positive weights.

Let the weights $a_{i}$ are all equal and denote $\bar \theta$ as the average contact duration, $\bar \theta = T_\text{meet}/K$, then (\ref{eq6}) become
\begin{equation}
 \frac{2}{\pi} \cdot K\cdot \sin \left( {\frac{\pi }{2} \cdot \frac{{\bar \theta}}{{{T}}}} \right) \ge \frac{2}{\pi} \cdot \sum\limits_{k = 1}^K {\sin \left( {\frac{\pi }{2} \cdot \frac{{\theta_k}}{T}} \right)}.
 \label{eq7}
\end{equation}
Thus, for two encounter patterns with the same frequency and duration, the more regular one gets a larger SRS value.
\end{IEEEproof}

\begin{Corollary}\label{Corollary}
The range space of the SRS value is $[0,1]$.
\end{Corollary}

According to the lemma 1-3, the SRS metric gets the maximum value when $T_\text{meet}$ is close to $T$ and each duration $t_k$ is almost the same and infinitesimal, and gets the minimum value when no encounter occurs. The following proof is about $\max \left(SRS\right)$.

\begin{IEEEproof}
The SRS metric gets the maximum when the following conditions hold:
\begin{itemize}
\item $\forall \theta_k, \theta_t \in \Theta_{v_i,v_j}, \theta_k = \theta_t$;

\item $T_{meet} = T$;

\item $K \rightarrow \infty$.
\end{itemize}

\begin{subequations}\label{eq8}
Since $\theta_k = \theta_t, \forall\theta_k, \theta_t \in \Theta_{v_i,v_j}$, the SRS metric can be derived as
\begin{equation}
{SRS_{v_i,v_j}} = \frac{2}{\pi} \cdot K \cdot \sin \left( \frac{\pi}{2} \cdot \frac{\bar \theta}{T}\right).
\end{equation}
Then $T_{meet} = T$, thus
\begin{equation}
{SRS_{v_i,v_j}} = \frac{2}{\pi} \cdot K \cdot \sin \left( \frac{\pi}{2} \cdot \frac{1}{K}\right).
\end{equation}
Finally, as $K\rightarrow \infty $, we have
\begin{equation}
{SRS_{v_i,v_j}} \rightarrow \frac{2}{\pi} \cdot \frac{\pi }{2} = 1.
\end{equation}
\end{subequations}
Thus $\max(SRS) = 1$.
\end{IEEEproof}
It is immediate to prove $\min \left(SRS\right) = 0$.
To illustrate the efficacy of the proposed SRS metric, we utilize the metric to evaluate the direct social relations of cases in Fig. \ref{fig1} and compare with existing methods, \textit{LocalCom} \cite{Li2009LocalCom} and \textit{TCCB} \cite{Zhou2017Predicting}. The experimental results  in Table \ref{tab1} show that  LocalCom and TCCB, which are based on $ASP$, fail to differentiate among cases (c)-(f), while our metric can accurately indicate which case supplies more interaction opportunities as argued earlier.

\begin{table}[ht]
\caption{Evaluation Results of the Cases in Fig. \ref{fig1}}
\begin{center}
\begin{tabular}{|c|c|c|c|c|c|c|}
\hline
\textbf{Cases} & a & b & c & d & e & f\\
\hline
\textbf{LocalCom} &0.483 & 0.485& 0.493 & 0.493 & 0.493 & 0.493\\
\hline
\textbf{TCCB} & 2.400 & 2.418 & 2.510 & 2.510 & 2.510 & 2.510\\
\hline
\textbf{SRS metric} & 0.199 & 0.487 & 0.494 & 0.716& 0.167 & 0.493\\
\hline
\end{tabular}
\label{tab1}\vspace{-.75cm}
\end{center}
\end{table}

\subsection{Analysis of Indirect Social Relations}\label{section3B}

According to the SRS metric, the direct influence between each pair of adjacent nodes can be evaluated. However, apart from direct influence, indirect influence among nodes may come from indirect interactions via neighboring nodes. This type of indirect influence also plays a significant role in centrality evaluation, especially in the absence of strong direct relations among nodes. To further explore indirect influence, an indirect SRS (\textit{in-SRS}) metric is proposed to reflect the indirect social relations among nodes. Considering a pair of nodes, $v_i$ and $v_j$, we say indirect influence exists between them if there is a set of nodes $Q=\{q_j\}$, $q_j \in V$ such that an indirect interaction between $v_i$ and $v_j$ can be bridged through those nodes. Here, the \textit{in-SRS} value is defined as the probability of influencing through all possible indirect interactions. Thus, the \textit{in-SRS} metric between $v_i$ and $v_j$ is expressed as
\begin{align}
in\text{-}SRS_{v_i,v_j} & = \Pr \left(l_1 \cup l_2 \cup \cdots \cup l_R\right) \\
& = 1 - \prod _{r = 1}^R {\left( {1 - PI_{v_i,v_j}\left(l_r\right)} \right)},
\label{eq9}
\end{align}
where $PI_{v_i,v_j}\left(l_r\right)$ denotes the indirect influence of the $r$-th indirect interaction, and is defined as the product of all intermediate direct social relations. Formally,
\begin{equation}
PI_{v_i,v_j}\left(l_r\right) = SRS_{v_i,q_{1}} \cdot \prod_{s = 1}^{S - 1} {SR{S_{{q_{s}},{q_{s+1}}}} \cdot } SR{S_{q_{S,v_j}}}.
\label{eq10}
\end{equation}

\subsection{Evaluation of Node Centrality}\label{section3C}

By evaluating the direct and indirect social relations, we can derive the influence range of each node and the influence strengths of it on the nodes within its influence range. We construct the influence sphere of each node by the set of nodes having the influence on it (including the direct and indirect). Formally, $IC_{v_i} = \{F,W\}$ denotes the influence sphere of node $v_i$, where $F$ is the friends set of $v_i$,
\begin{equation}
	\begin{aligned}
		F\left(v_i\right) = \{ v_j|&SRS(v_i,v_j) \neq 0 \\
		&\mbox{ or }in\text{-}SRS(v_i,v_j) \neq 0\},
	\end{aligned}
\label{eq11}
\end{equation}
and $W \left(v_i\right) = \{w_{v_i,v_j}\}$, $v_j \in F$, is a set of influence strengths between $v_i$ and its friends,
\begin{equation}
w_{v_i,v_j} = 1 - \left(1 - SRS_{v_i,v_j}\right) \left(1 - in\text{-}SRS_{v_i,v_j}\right).
\label{eq12}
\end{equation}

Since each influence sphere contains all possible influence members of a node, the task of quantifying the node centrality in the whole networks shifts to quantifying the influence of nodes in their influence spheres. Hence, this paper proposes a Social-Relation based Centrality (SoReC) measure to quantify the centrality of users on the basis of Entropy theory.

The entropy notion is introduced in thermodynamics and has been widely used in information science and statistical physics to describe the probability distribution of a given system. In this paper, we employ entropy to evaluate the distribution of influence strengths in the influence spheres. Consider a influence sphere of $v_i$, $IC_{v_i}$. The influence probability of node $v_i$ on node $v_j \in IC_{v_i}$ is expressed as
\begin{equation}
P_{v_i}\left(v_j\right) = \frac{w_{v_i,v_j}}{\sum_{v_q \in IC_{v_i}}w_{v_i,v_q}}.
\label{eq13}
\end{equation}
Thus, the influence entropy of node $v_i$ is defined as
\begin{equation}
 H\left( v_i \right) =  - \sum_{v_j \in IC_{v_i}} {{P_{v_i}}(v_j) \cdot {{\log }_2}{P_{v_i}}(v_j)}.
 \label{14}
\end{equation}

Based on Entropy theory, the node with wider influence range and uniform influence probability has higher influence entropy. However, the influence entropy slights the weights of social relations, which reflect the actual influence among nodes. Thus we add the weight information into centrality measure and update the measure as
\begin{equation}
  SoReC\left( v_i \right) = H\left( v_i \right) \cdot \sum\limits_{q_i \in IC_{v_i}} {w_{v_i,q_i}}.
 \label{15}
\end{equation}

\begin{figure*}[!t]
\centering
\subfloat[Influence range]{\includegraphics[width=0.295\textwidth]{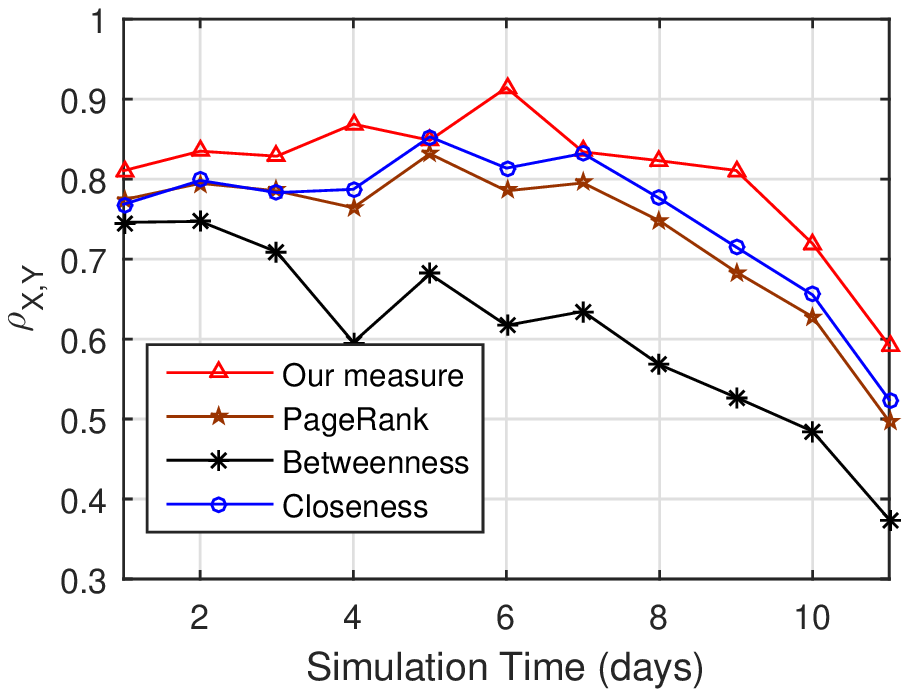}
\label{fig2casea}}
\hfil
\subfloat[Influence speed]{\includegraphics[width=0.295\textwidth]{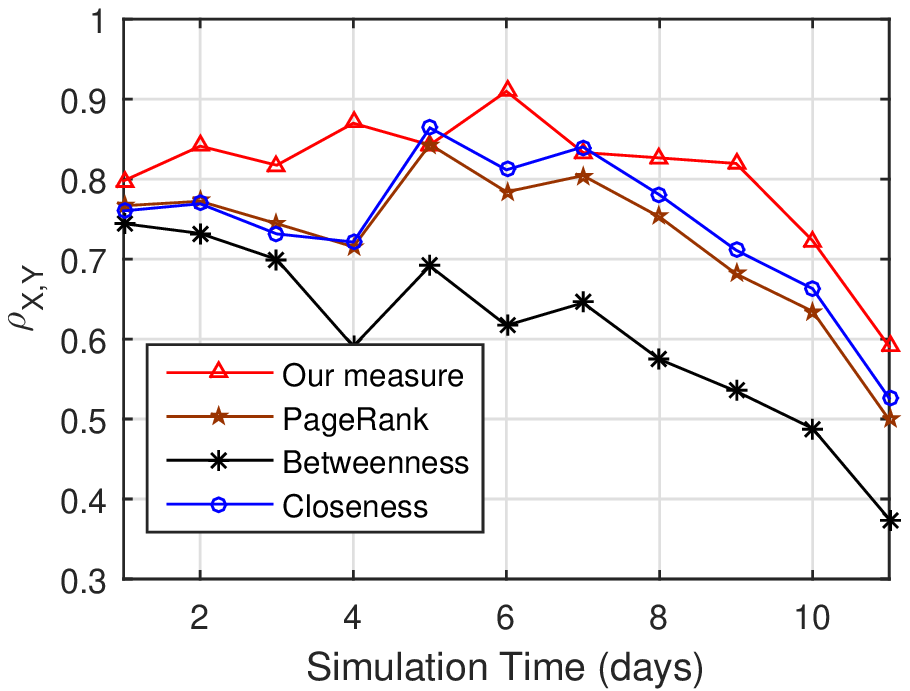}
\label{fig2caseb}}
\hfil
\subfloat[Accuracy]{\includegraphics[width=0.3\textwidth]{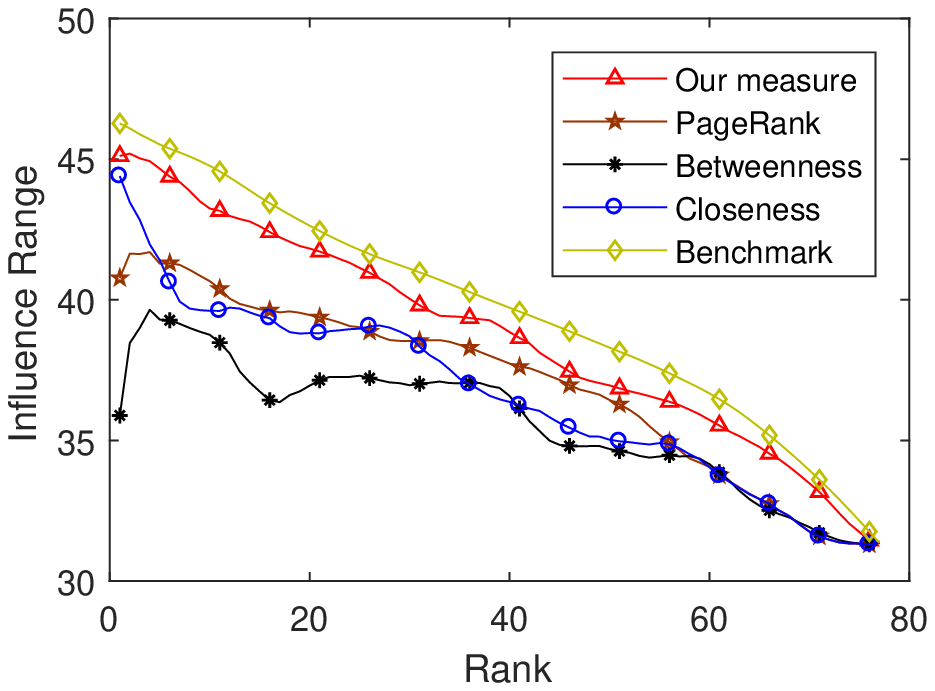}
\label{fig2casec}}
\hfil\vspace{-.45cm}
\subfloat[Influence range]{\includegraphics[width=0.295\textwidth]{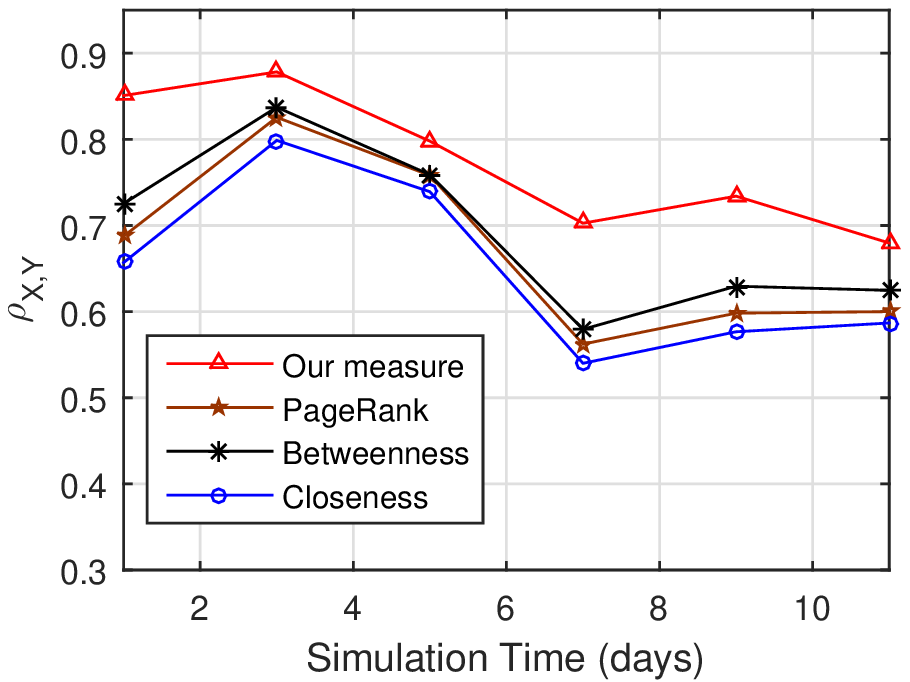}
\label{fig2cased}}
\hfil
\subfloat[Influence speed]{\includegraphics[width=0.295\textwidth]{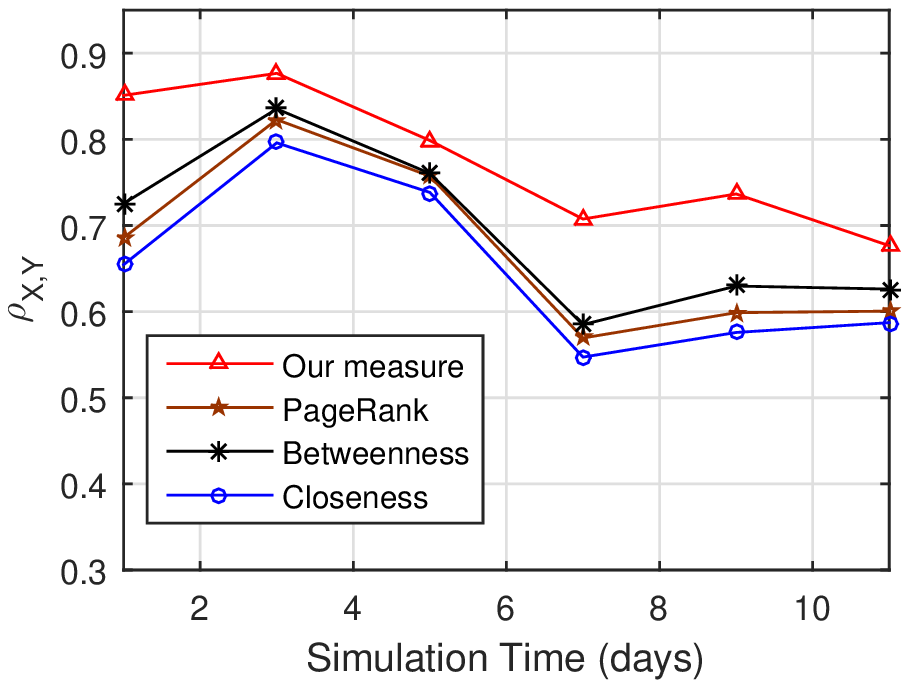}
\label{fig2casee}}
\hfil
\subfloat[Accuracy]{\includegraphics[width=0.3\textwidth]{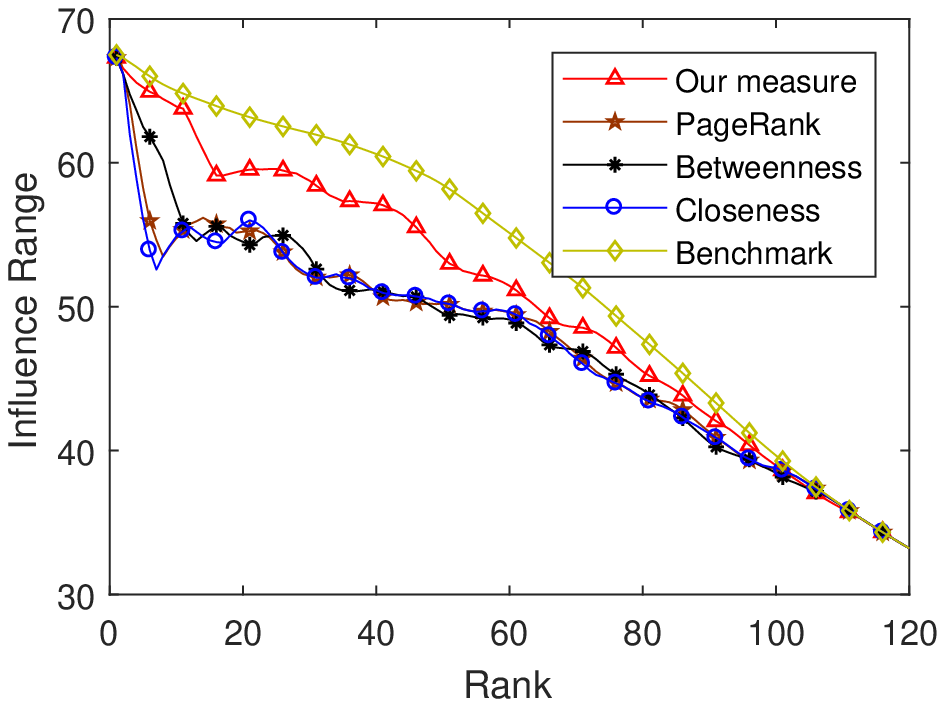}
\label{fig2casef}}
\caption{The simulation results under different indicators, where (a)-(c) are the Pearson correlation coefficients between predicted centrality rank with actual influence range rank and actual influence speed rank as well as the measurement accuracy under the Reality traces, while (d)-(f) are those under the WTD traces.}
\label{fig2}\vspace{-.5cm}
\end{figure*}

\section{Simulation}\label{section4}

Next, we employ the Susceptible-Infected-Recovered (SIR) model \cite{Newman2002Spread} to simulate the spreading process on MSNs and carry the simulations on the real-world mobility networks.

\subsection{Experimental Settings}\label{section4a}

\subsubsection{Dataset}

The simulations in this work are based on two widely used real-world datasets: \textit{(i)} MIT Reality Mining Data (Reality) \cite{Eagle2009Inferring}, and \textit{(ii)} UCSD Wireless Topology Discovery Trace (WTD) \cite{Mcnett2003Access}. In Reality, Bluetooth data are recorded by 97 smartphones deployed on students and staff at MIT over 246 days. In WTD, WiFi data are recorded by 275 PDAs carried by freshmen students at UCSD over 11 weeks. The details about the datasets are illustrated in Table \ref{tab2}. We extract the contact records from the partial data of the two datasets for our simulations. Each processed records includes the start and end time of each encounter and the IDs of the nodes in contact.

\subsubsection{SIR Model}
We use the SIR model to simulate spreading processes on networks and test the influence of every node. In SIR, every node is initialized to be the \textit{susceptible} state, and they may convert to the \textit{infected} state with probability $\lambda$ when contacting an infected node. In addition, the infected nodes may recover over time, and recovered nodes will not be infected. During the simulation period, for a given initial infected node, the number of infected and recovered nodes (influence range), and the average time of infection (influence speed) are recorded and used as its actual influence ability.

\begin{table}[t]
\caption{Characteristics of Two Datasets}
\begin{center}
\begin{tabular}{|c|c|c|}
\hline
\textbf{Dataset} & Reality \cite{Eagle2009Inferring} & WTD \cite{Mcnett2003Access} \\
\hline
\textbf{Device} & Phone & PAD\\
\hline
\textbf{Network type} & Bluetooth & WiFi\\
\hline
\textbf{Contact type} & direct & Ap-based\\
\hline
\textbf{Duration (days)} & 246 & 77\\
\hline
\textbf{Number of nodes} & 97 & 275\\
\hline
\textbf{Number of contacts} & 54,667 & 135,364\\
\hline
\end{tabular}
\label{tab2}\vspace{-.75cm}
\end{center}
\end{table}

\subsubsection{Evaluation Metrics}

Pearson correlation coefficient is used to test whether the influence range (or the influence speed) correlates with nodes centrality values under different situations, which can be expressed as
\begin{equation}
\rho_{X,Y} = Cor\left(X,Y\right),
\label{16}
\end{equation}
where $X$ is the ranking list by different centrality measures and $Y$ is th{}e ranking list by the actual influence ability.

\subsection{Experimental Results}\label{section4b}

Firstly, we evaluate the prediction ability of the proposed measure. In this pa{}rt, the former portion of the dataset is used as the contact history data for the centrality quantification. The remaining portion is used as the test data for the actual influence ability test. In addition, we also evaluate the performance of traditional measures, including Betweenness, Closeness, \cite{Freeman1978Centrality}, and PageRank \cite{Sehgal2009The}, as baseline comparison.

Figs. \ref{fig2}a, \ref{fig2}b, \ref{fig2}d, \ref{fig2}e illustrate the correlation between the actual influence strength (range and speed) and the predicted value of node centrality under different networks. The performance of each measure varies in different networks. In Reality, the Closeness measure performs better than the Betweenness, but the opposite is true in WTD. But overall, the four measures can well predict the centrality of nodes in the future and the performance decreases with the increasing period between prediction and test. By contrast, the correlation coefficients of our measure are systematically the largest, which means that the SoReC measure can quantify the centrality of nodes more accurately in dynamic MSNs.
In addition, we detail the ability of our metric in the centrality evaluation, shown in Fig. \ref{fig2}d, \ref{fig2}f. Since the effects turned out to be similar on influence range and speed, we only show the results on the influence range. In Fig. \ref{fig2}d, \ref{fig2}f, the X-axis is the rank of nodes and Y-axis is the average actual influence range of top-L nodes ranked by different centrality measures. Notice that the benchmark curve is based on the actual rank, i.e., the benchmark list is ranked by the actual influence range. We can observe that our curves are closest to benchmark curve under both datasets, and the advantage of our method is most marked in the head of the distributions. The results illustrate that our measure has an advantage in the centrality evaluation over other methods, especially in the identification of influential users.

\section{Conclusions}\label{section5}
In this paper, we captured social relations to study links among users, and on this basis, proposed  \textit{SoReC} to identify influential users in MSNs. Through theoretical derivations and experimental verification, the SoReC measure we proposed is proved to able to accurately quantify the centrality of nodes in MSNs. In addition, the SoReC measure performs better than traditional measures in terms of centrality prediction. Despite the promising results, our model still requires a knowledge of the global network topology. In the future, we will attempt to identify the influential users in a distributed fashion, or by relying on a mix of global and local information \cite{de2014mixing}. We will then leverage our framework to redesign proofs-of-concept of some popular services and applications in MSNs.

\balance
\bibliographystyle{IEEEtran}
\bibliography{MSNs1}

\end{document}